\documentclass[useAMS]{mn2e}
\usepackage{amsmath}
\usepackage{times}
\usepackage{psfig}
\usepackage{graphics,graphicx}
\usepackage{lscape}
\usepackage{epsf}
\usepackage{epsfig}
\usepackage{color}
\usepackage{dcolumn}

\newcommand{\xmm}{{\em XMM-Newton}}

\newcommand{\tsco}{\mbox{$\tau$\,Sco}}

\newcommand{\Lx}{\mbox{$L_{\rm X}$}}

\def \etal   {\hbox{et~al.\/}}
\title[X-ray Study of $\tau$ Sco Like Stars]
{A Report on the X-ray Properties of the $\tau$~Sco Like Stars}

\author[Ignace, Oskinova, and Massa]
{R.~Ignace$^{1}$\thanks{E-mail:  ignace@etsu.edu}, L.~M.~Oskinova$^{2}$,
and D.~Massa$^{3}$ \\
$^{1}$Physics \& Astronomy, East Tennessee State University, Johnson City,
	TN, USA \\
$^{2}$ Institute for Physics and Astronomy, University of Potsdam,
14476 Potsdam, Germany \\
$^{3}$ Space Telescope Science Institute, 3700 San Martin Drive,
Baltimore, MD 21218, USA}

\begin{document}
\maketitle

\label{firstpage}

\begin{abstract}

An increasing number of OB stars have been shown to possess magnetic
fields.  Although the sample remains small, it is surprising that
the magnetic and X-ray properties of these stars appear to be far
less correlated than expected.  This contradicts model predictions,
which generally indicate that the X-rays from magnetic stars to be harder
and more luminous than their non-magnetic counterparts.   Instead,
the X-ray properties of magnetic OB stars are quite diverse.

$\tau$~Sco is one example where the expectations are better met.
This bright main sequence, early B star has been studied extensively
in a variety of wavebands.  It has a surface magnetic field of
around 500~G, and Zeeman Doppler tomography has revealed an unusual
field configuration.  Furthermore, $\tau$~Sco displays an unusually
hard X-ray spectrum, much harder than similar, non-magnetic OB
stars.  In addition, the profiles of its UV P~Cygni wind lines have
long been known to possess a peculiar morphology.

Recently, two stars, HD~66665 and HD~63425, whose spectral types
and UV wind line profiles are similar to those of $\tau$~Sco, have
also been determined to be magnetic.  In the hope of establishing
a magnetic field -- X-ray connection for at least a sub-set of the
magnetic stars, we obtained XMM-Newton EPIC spectra of these two
objects.  Our results for HD~66665 are somewhat inconclusive.  No
especially strong hard component is detected; however, the number
of source counts is insufficient to {\em rule out} hard emission.
longer exposure is needed to assess the nature of the X-rays from
this star.  On the other hand, we do find that HD~63425 has a
substantial hard X-ray component, thereby bolstering its close
similarity to $\tau$~Sco.  

\end{abstract}

\begin{keywords}
stars:  early-type; stars: individual: HD63425, HD66665;
stars:  magnetic field; X-rays:  stars
\end{keywords}

\section{Introduction}

In spite of their apparent simplicity, near main sequence B stars
exhibit a range of properties that are not well understood.  Among
the challenges include: surprisingly low wind mass-loss rates and
wind terminal speeds (Petit \etal\ 2011; Oskinova \etal\ 2011), a
full understanding of the causes and evolution of the Be phenomenon
(e.g., Porter 1999; Brown, Cassinelli, \& Maheswaran 2008;
Townsend, Owocki, \& Howarth 2004; Carciofi
\etal\ 2009; Wisniewski \etal\ 2010), and the nature of magnetism
being detected among some B stars (Hubrig \etal\ 2006; Rivinius
\etal\ 2010; Petit \etal\ 2011; Oksala \etal\ 2012; Grunhut \etal\
2012).

In regard to their mass loss, a number of main sequence B stars
display mass-loss rates $\dot{M}$ that are an order of magnitude
lower than theoretical expectations.  Examples include detailed
spectral analyses of five B stars described in Oskinova \etal\
(2011):  $\tau$~Sco, $\beta$~Cep, $\xi^1$~CMa, V2052~Oph, and
$\zeta$~Cas.  Although $\beta$~Cep is a giant, the other four are
B0--2 main sequence stars.  In addition, the early BV stars HD~66665
and HD~63425, which are the subject of this paper, were analyzed
by Petit \etal\ (2011) and found to have low $\dot{M}$ values, an
order magnitude lower than predicted by Vink, de Koter, \& Lamers
(2000).  Notably, the presence of X-rays in their winds played an
important, if not central, role in achieving satisfactory fits to
observed UV and optical spectra.  It seems then that at least some
B stars exhibit the same ``weak wind'' problem seen in the less
luminous O~stars (e.g., Martins \etal\ 2005; Marcolino \etal\ 2009;
Muijres \etal\ 2012; Lucy 2012; Huenemoerder
\etal\ 2012).

Then there is the occurrence, properties, and evolutionary
influence of magnetic
fields among B stars.  Our contribution to this issue has been to
study the X-ray emissions from magnetic B stars in order to identify
relationships (or the absence thereof) between known magnetic
properties and measured X-ray characteristics (Ignace \etal\
2010; Oskinova \etal\ 2011).

Our attempt to draw connections between magnetism and X-rays among
the B~stars was inspired by several by successes in relating
X-ray properties to magnetospheric models for early-type stars with
strong magnetic fields.  Magneto-hydrodynamical simulations of the
early O~star $\theta$~Ori$^1$ appear capable of matching the observed
X-ray variations, both in broad band terms as well as in emission
lines (Gagne \etal\ 2005; ud-Doula 2012).  The Bp star $\sigma$~Ori~E
has a very strong surface magnetic field of $\approx 20,000$~G that
motivated a semi-analytic approach called the Rigidly Rotating
Magnetosphere (RRM) model (Townsend \& Owocki 2005).  The RRM has
been broadly successful in explaining observed H$\alpha$ variations,
the polarization light curve, and with extension to time-dependent
hydrodynamics, the star's broad X-ray properties (Townsend, Owocki,
\& ud-Doula 2007).  Indeed, the approach has even been able to
explain the {\em measured} rotational spin-down rate of the star
(Townsend \etal\ 2010).  Although MHD simulations still face
challenges with producing detailed quantitative matches to
observed X-rays of magnetic massive stars (e.g., Naz\`{e} \etal\
2010), the models are a work in progress that offer a promising framework in
which to interpret the observations.

With this framework, it was thought that deeper insights
into the relationship between magnetic and X-ray characteristics
could be gained through a study of 
B~stars with weaker yet
moderately strong surface magnetic fields and lower wind mass-loss
rates.  To this end, the
efforts of the MiMeS (e.g., Grunhut \& Wade 2012) and Magori
(Scholler \etal\ 2011) collaborations to detect, characterize, and
catalog the magnetic properties of early-type stars have been
indispensable.  Unfortunately, a clear relation between stellar
magnetism and X-ray fluxes has not emerged.  Indeed, the apparent
{\em absence} of expected relationships between magnetic and X-ray
properties has been a surprise (Favata \etal\ 2009; Ignace \etal\
2010; Oskinova \etal\ 2011).

Despite the lack of an overall connection, it may be that certain
types of magnetic stars do exhibit one.  For example, $\tau$~Sco
is a magnetic star with unusual UV wind lines, and is also 
notable for having an unusually hard component to its X-ray
emission for a massive star, especially for an early B~type
star that is believed to be single (e.g., Cassinelli \etal\ 1994;
Cohen \etal\ 2003; Mewe \etal\ 2003; Ignace \etal\ 2010).
Recently, two other stars, HD~66665 and HD~63425, were identified
as having UV wind lines with the same peculiar morphology seen in
$\tau$~Sco.  This motivated Petit \etal\ (2011) to observe both
stars, and both yielded significant positive detections of surface
magnetic fields.  We refer to Petit \etal\ (2011) for a discussion
and spectral analysis of HD~66665 and HD~63425, and to Oskinova
\etal\ (2011) for a spectral analysis of $\tau$~Sco.

The question that naturally arises is whether or not HD~66665 and
HD~63425 are also hard sources of X-rays like $\tau$~Sco. If so,
the discovery would produce a rare example among massive stars of
a relationship involving stellar magnetism, X-ray emissions, and
UV line profile morphology.
We report here on data obtained with the {\em XMM-Newton}
in an effort to characterize the X-ray luminosities and hot plasma
temperatures for HD~66665 and HD~63425.  Section~\ref{sec:obs}
details the acquisition and reduction of data obtained with the
EPIC detectors.  Section~\ref{sec:xspec} presents an analysis of
the X-ray spectra.  An assessment of whether HD~66665 and HD~63425
are indeed hard sources is given in Section~\ref{sec:results},
followed by concluding remarks in
section~\ref{sec:conc}.

\begin{table}
\begin{center}
\caption{Properties of Analogue Stars$^\dag$\label{tab1}}
\begin{tabular}{lcc}
\hline\hline
 & HD 63425 & HD 66665 \\ \hline
Type & B0.5V & B0.5V \\
$T_{\rm eff}$ (K) & 29,500 & 28,500 \\
$\log g$ (cgs) & 4.0 & 3.9 \\
$\log L_\ast/L_\odot$ & 4.50 & 4.25 \\
$M_\ast /M_\odot$ & 17 & 9 \\
$R_\ast/R_\odot$ & 6.8 & 5.5 \\
$v \sin i$ (km/s) & $< 15$ & $<10$ \\
$\dot{M}$ ($M_\odot$/yr) & $<7.5 \times 10^{-10}$ & $<4.5 \times 10^{-10}$ \\
$v_\infty$ (km/s) & $\sim 1700$ & $\sim 1400$ \\
$d$ (kpc) & 1,136 & 1--2 \\
$\log N_H$ (cm$^{-2}$) & 20.64 & 20.15 \\
$V$ magn & 6.9 & 7.8 \\ \hline
\end{tabular}

$^\dag$ Values taken from Petit \etal (2011), except for
the hydrogen column density that comes from Diplas \& Savage
(1994)
\end{center}
\end{table}

\section{Observations and Data Reduction	\label{sec:obs}}

\label{sec:obs}

We obtained dedicated \xmm\ observations of HD\,63425\ and HD\,66665.
Stellar and wind properties of our target stars are given
in Table~\ref{tab2}.
All three (MOS1, MOS2, and PN) European Photon Imaging Cameras
(EPICs) were operated in the standard, full-frame mode and a medium
UV filter.  A log of observations is shown in Table\,\ref{tab2}.
The data were analyzed using the software {\sc sas}\,10.0. The time
periods when the particle background was high were excluded from
the analysis. Both stars were detected by the standard source
detection software.  The exposure times and EPIC PN count rates for
our program stars are given in Table\,\ref{tab2}.

A bright patch of diffuse X-ray emission with diameter of $\approx
4$\,arcmin is present in the EPIC images of HD\,63425.  The spectrum
of the diffuse emission was found to be well fitted with a two
temperature plasma having components $kT_1\approx 0.7$\,keV and
$kT_2\approx 5.4$\,keV. The X-ray temperature, flux, brightness
distribution, and comparison with optical and IR images indicate
that this diffuse emission is most likely due to a massive galaxy
cluster at $z>0.3$ (A.~Finoguenov, private comm.).  The spectrum
of HD\,63425\ was extracted from a region with a diameter of $\approx
15''$. The X-ray background was chosen from a nearby area in the
diffuse X-ray source. Thus, it is possible that the hard stellar
X-ray emission for HD~63425 is {\em over-subtracted} because of the
hard background diffuse radiation. Therefore, the X-ray spectrum
of HD\,63425\ presented here provides only a conservative estimate
of the hottest temperature plasma component.

The X-ray point source with the coordinates of HD\,66665\ is well
isolated, and there was no difficulty in obtaining its spectrum using
the standard procedure and determining the X-ray background from a nearby
region free of X-ray sources.

\begin{table}
\begin{center}
\caption{\xmm\ Observations of \tsco-Analogue Stars	\label{tab2}}
\vspace{1em}
\renewcommand{\arraystretch}{1.2}
\begin{tabular}[h]{lccl}  \hline
\hline Star & MJD   &  useful exposure & PN count-rate$^{\rm a}$ \\
            &       &  [ksec]         & [s$^{-1}$] \\
\hline
HD\,63425\ & 55687.8474 & 12  & $0.064 \pm\ 0.002$     \\
HD\,66665\ & 55077.0906 & 25  & $0.022 \pm\ 0.001$ \\
\hline
\end{tabular}

{\small $^{\rm a}$ In the 0.3-7.0\,keV band; background subtracted.}
\end{center}
\end{table}

\section{Results}
\label{sec:xspec}

To analyze the spectra we used the standard spectral fitting software
{\sc xspec} (Arnaud 1996).  The number of counts per bin in
the spectra of HD\,63425\ and HD\,66665\ is small; therefore, we
used the Cash-statistic (Cash 1979) for spectral fitting.
Using the neutral hydrogen column density as a fitting parameter
does not yield a sensible constraint on its value; therefore, $N_{\rm
H}$ was fixed at its interstellar value (see Tab.~\ref{tab1}).

Our targets are known magnetic stars, and peculiar abundances are
often found in such stars, typically explained as arising from
diffusion processes which allow heavier elements to sink in the
atmosphere under the influence of gravity, while lighter elements
are elevated to the surface by radiation pressure (e.g., Morel
\etal\ 2008).  It is usual for a magnetic star to show an overabundance
of nitrogen, and sometimes helium.  For example, Morel (2011) find
that the abundance ratio [N/C] is higher than solar in HD\,66665,
while [N/O] is nearly solar. We are not aware of any abundance
studies for HD\,63425. The quality of the X-ray spectra of our
program stars are not sufficient to constrain abundances.  We carried
out tests that showed that the overabundance of N by a factor of a
few does not significantly change the results of our spectral fits.
Therefore, abundances for our two target stars were set  to solar
values based on Asplund (2009).

\subsection{HD\,63425} 

\begin{figure}
\centering
\includegraphics[height=0.99\columnwidth, angle=-90]{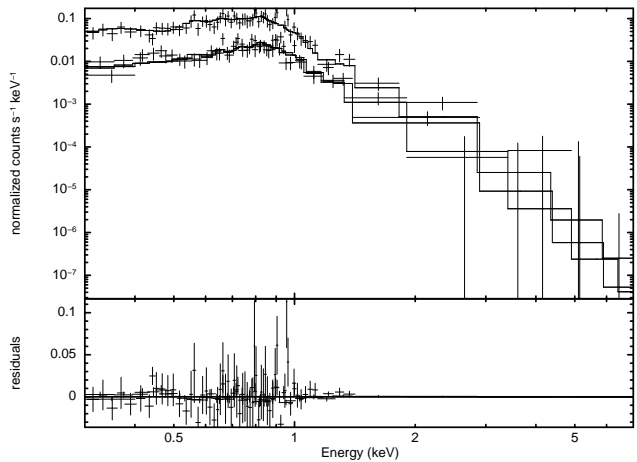}
\caption{ \xmm\ PN (upper curve), and MOS1 and MOS2 (lower curves)
spectra of HD\,63425\ with the best fit
three-temperature model (solid lines). The model parameters are
shown in Table\,\ref{tab3}.}
\label{fig:hd63425}
\end{figure}

Our \xmm\ observation detected the X-ray emission from HD\,63425\
for the first time.  The 90\% confidence range for the unabsorbed
X-ray flux, meaning the intrinsic flux of the star after correcting for
interstellar absorption, is $1.31-1.71 \times 10^{-13}$ erg s$^{-1}$
cm$^{-2}$.
Assuming a distance of $d=1.136$\,kpc, the X-ray
luminosity of  HD\,63425\ is $\Lx\approx 2\times 10^{31}$\,erg\,s$^{-1}$
with an error of about 15\%;
this X-ray luminosity is comparable to the value for \tsco.  The observed
EPIC spectra of HD\,63425\ and the fitted model are shown in
Figure~\ref{fig:hd63425}.  A two temperature plasma model can
reproduce the observed spectrum quite well (see Tab.~\ref{tab3}).

\subsection{HD\,66665}  

\begin{figure} 
\centering \includegraphics[height=0.99\columnwidth,angle=-90]{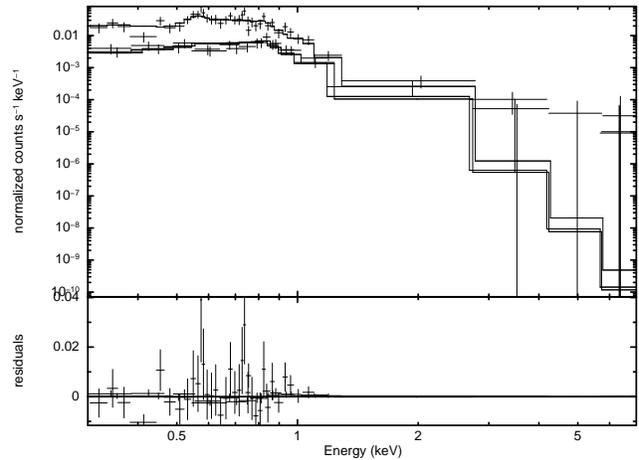} 
\caption{ \xmm\ EPIC-PN spectrum of HD\,66665\ and the best fit
two-temperature model. The model parameters are shown in
Table\,\ref{tab3}.} \label{fig:hd66665} 
\end{figure}

Our \xmm\ observation detected the X-ray emission from HD\,66665\
also for the first time. The source has only a modest count rate (see
Tab.~\ref{tab2}).  The unabsorbed X-ray flux is $2.0-4.8 \times 10^{-14}$
erg s$^{-1}$ cm$^{-2}$.
Assuming a distance of $d=1-2$\,kpc, the X-ray
luminosity of HD\,66665\ falls in the conservative range
of $\Lx\approx 2-22\times 10^{30}$\,erg\,s
$^{-1}$.  
The
EPIC spectra of HD\,66665\ can be well described using a two
temperature plasma model (see Tab.~\ref{tab3}).  The observed spectra
and a model fit are shown in Figure~\ref{fig:hd66665}.  It is
interesting to note that the emission measures of hotter and cooler
plasma components are quite similar. This is in contrast to other
magnetic B-type stars, where the softer component usually has much
larger emission measure (c.f., Oskinova \etal\ 2011); however,
$\tau$~Sco is one notable exception to this rule.

Although the two-temmperature fit is statistically acceptable, it
seems that the model doesn't reproduce well the spectral shape at
energies above 2\,keV. We attempted to find a three-temperature
model fit or a power-law fit, but these additional model components
were essentially unconstrained.  Thus, while it appears that there
are indications of a harder component being present in the spectrum
of HD\,66665, it must be confirmed by better quality data.

\begin{figure*}
\centerline{\epsfig{file=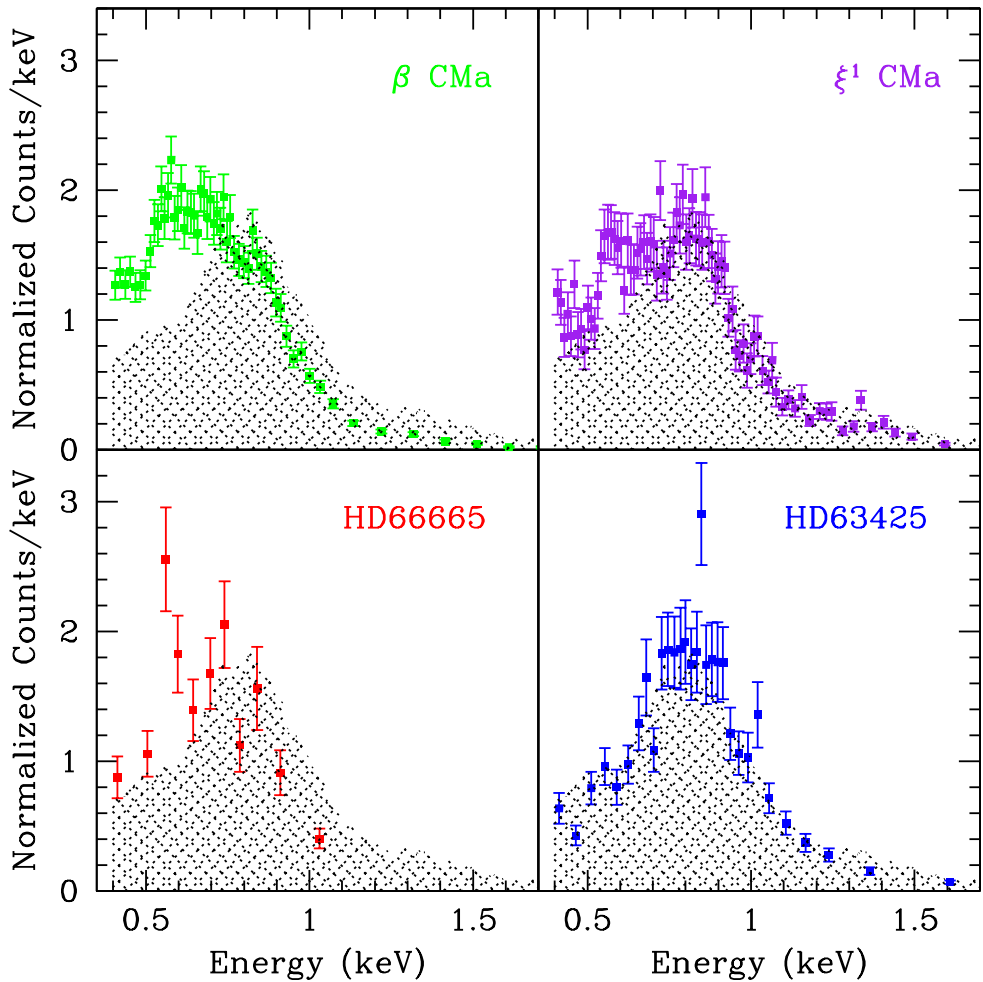,angle=0,width=14cm}}
\caption{
A comparison of XMM-Newton spectra of five B stars from the PN
instrument, as labeled.  Only $\beta$~CMa is known not to be
magnetic; all the others are magnetic stars.  For reference 
the spectrum of $\tau$~Sco is displayed as the hatched area in
each figure.  All of the spectra have been normalized to unit area
for comparisons of spectral shape.
\label{fig:comparisons}
}
\end{figure*}

\section{Discussion}	\label{sec:results}

\subsection{Comparison of Spectra}

Figure~\ref{fig:comparisons} displays a comparison of the X-ray
spectra of the two $\tau$~Sco analogues against $\tau$~Sco
itself, as well as two other reference objects, $\xi^1$~CMa and
$\beta$~CMa.  Each source spectrum has been normalized to unit area
for the sake of comparison.  The spectrum of $\tau$~Sco is shown
as the hatched region in each panel of this figure.  The two analogue
objects are shown at bottom; the other two reference objects at
top.  The B~star $\xi^1$~CMa is a magnetic star with a surface field
of about 1,450~G (Hubrig \etal\ 2006; Fortune-Ravard \etal\ 2011);
its X-ray properties have been reported in Oskinova \etal\ (2011).
The source $\beta$~CMa is a giant B~star that does not, so far,
have a detectable magnetic field (Hubrig 2006).  The star has been
observed with the XMM-EPIC (PI: W.~Waldron), 
but a detailed analysis has not been
reported in the literature.  Here we present only a preliminary
spectrum of $\beta$~CMa for the purpose of having a high signal-to-noise
X-ray spectrum with (a) the same instrument as our analogues sources
and (b) which is known not to have a significant surface magnetic
field.

Normalization of the spectrum accentuates differences in the spectral
energy distributions between the repective sources and $\tau$~Sco.
(Note:  With an EPIC/PN spectrum of over 100,000 X-ray counts, the
S/N of $\tau$~Sco's spectrum is so much higher than the other
stars that we do not show error
bars.)  The spectra of both HD~63425 and $\xi^1$~CMa closely hug
the shape of $\tau$~Sco's spectrum.  By contrast both HD~66665
and $\beta$~CMa show peak values that are shifted to softer energies
and a relative deficit of quite hot gas as compared to $\tau$~Sco.

There are two main comments to be made at this point.  First,
$\beta$~CMa is at an extremely low interstellar hydrogen column
density, approximately two orders of magnitude lower than
the other four stars (see Tab.~\ref{tab3}).  In fact, because
of its low column density, $\beta$~CMa was one of only two massive
stars observed with the EUVE (Cassinelli \etal\ 1996).  For the present
analysis, the low column results in minimal attenuation of the softer
X-ray emissions from this star, which naturally shifts the
X-ray spectral peak of $\beta$~CMa to lower energies.  
Still, as will be discussed,
$\beta$~CMa lacks a substantial hard component to its X-ray spectrum.

The second point is that the overall counts for HD~66665 are low,
lowest of all five sources in this report.  The lower-than-expected
count rate of HD~66665 suggests that hard emission could be present
but not detected.  In effect, a low level of hard emission above
1.5~keV could be present intrinsically, but lost in the background
noise owing to insufficient counts.  As a result, only the dominant
softer component survives in the data reduction.  Our main
conclusion for HD~66665 is that its spectrum does not provide
evidence of hard emission, but that a longer exposure is needed to
determine confidently whether or not hard emission is produced by
the system.

\begin{table*}
\begin{center}
\caption{X-ray Characteristics of Sources$^a$	\label{tab3}}
\begin{tabular}{lccccc}
\hline\hline
 & $\tau$ Sco$^b$ & HD 63425 & HD 66665 & $\xi^1$ CMa & $\beta$ CMa \\ \hline
 & & & & & \\
$N_H^c$ ($10^{20}$ cm$^{-2}$) & 2.0 & 4.4 & 1.4 & 1.4 & 0.02 \\
PN Count Rate (cps)  & 7.99 & 0.063 & 0.023 & 0.67 & 0.73  \\
 & & & & & \\
$d$ (kpc) & 0.145 & 1.11 & 1--2 & 0.420 & 0.151 \\
$L_X$ ($10^{30}$ erg/s) & 32 & $12\pm 2$ & 2--23 & 31 & $3.2\pm 0.2$ \\
$L_X/L_\ast$ ($10^{-7}$) & 4.0 & 1.6 & $2-22$ & 2 & 0.4 \\
$B_\ast$ (G) & 500 & $\sim 1,100$ & 600 & 1,450 & 0 \\
 & & & & & \\
$kT_1$ (keV) & $0.141 \pm 0.005$ & $0.19 \pm 0.01$ & $0.16 \pm 0.04$ & $0.121\pm 0.004$ & $0.111\pm 0.004$\\
$EM_1/EM_T$ & $0.17\pm 0.02$ & $0.54\pm 0.10$ & $0.50\pm 0.31$ & $0.65\pm 0.14$ & $0.61\pm 0.12$ \\
 & & & & & \\
$kT_2$ (keV) & $0.85 \pm 0.06$ & $0.60 \pm 0.02$ & $0.39 \pm 0.06$ & $0.564\pm 0.009$ & $0.353\pm 0.007$ \\
$EM_2/EM_T$ & $0.83\pm 0.06$ & $0.46\pm 0.08$ & $0.50\pm 0.34$ & $0.35\pm 0.04$ & $0.39\pm 0.05$ \\ 
 & & & & & \\
$\langle kT \rangle$ (keV) & $0.73\pm 0.07$ & $0.38\pm 0.05$ & $0.28\pm 0.15$ & $0.28 \pm 0.03$ & $0.21\pm 0.02$ \\
$\Delta kT/\sigma_\Delta^d$ & 7.3 & 3.1 & 0.5 & 1.9 & N/A \\
 & & & & & \\\hline
\end{tabular}

{\small 
$^a$ Values based on fits to the EPIC-PN spectral data, unless indicated otherwise.\\
$^b$ Values based on RGS/MOS analysis by Mewe \etal\ (2003); see text. \\
$^c$ Values fixed as based on measures found in the literature. \\
$^d$ See text for an explanation of this ratio.}
\end{center}
\end{table*}

\subsection{Statistical Analysis}

With the exception of $\tau$~Sco, we have made two-temperature fits
to our sources.  For $\tau$~Sco, the quality of the spectrum is so
high, at over 100,000 counts detected, that a two-temperature fit
produces a poor match to the spectrum.  In this case we use the
four-temperature fit of Mewe \etal\ (2003) in the following discussion
of source X-ray properties.

X-ray spectral characteristics are given for the five stars
under discussion in Table~\ref{tab3}.  The
table lists the hydrogen column density, X-ray count rate in EPIC/PN,
the X-ray luminosity from EPIC/PN, and the temperatures (as $kT$ in keV)
and relative emission measures of the two temperature fits.  Also
listed is an emission-measure-weighted average temperature, defined
by

\begin{eqnarray}
\langle kT \rangle & = & \frac{\sum_{\rm i}\, kT_{\rm i}\,EM_{\rm i}}{\sum_{\rm i}\, EM_{\rm i}} \nonumber \\ 
 & = & \sum_{\rm i}\,\left(\frac{EM_{\rm i}}{EM_T}\right)\,kT_{\rm i}, \\
\end{eqnarray}

\noindent where $EM_T$ is the total emission measure.  In the case
of $\tau$~Sco, the star has one measure of temperature of relatively
low value, typical of other OB stars, and three higher temperature
components.  Those three higher ones have been emission-measure
averaged according to the values quoted by Mewe \etal\ (see their
Tab.~2), and given
in the table simply as $kT_2$.  

Table~\ref{tab3} also gives the measured surface magnetic field
values, the X-ray luminosities, and the ratios of X-ray to Bolometric
luminosity.  The surface fields show a large range.  Not counting
the non-magnetic star $\beta$~CMa, field values range from about
500~G for $\tau$~Sco to one that is $3\times$ that for $\xi^1$~CMa.
Although none of the stars are exceptional in their value of
$L_X/L_{\rm Bol}$, having ratios of order $10^{-7}$ that is
reflective of the 
standard found for other O stars and early B stars 
(e.g., Berghoefer \etal\ 1997), it is interesting to note that the
ratio for $\beta$~CMa is smaller than all of the magnetic stars
being considered.

\begin{figure}
\centerline{\epsfig{file=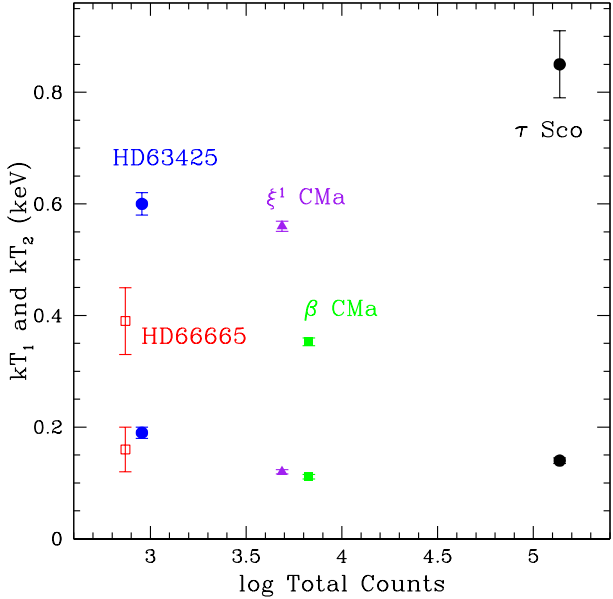,angle=0,width=8.5cm}}
\caption{
A plot of the two temperature components from spectral fits,
provided as $kT$ in units of keV, versus the total counts
for the source.  For $\tau$~Sco, the higher temperature component
is an emission measure  weighted average from the three hot components described in
Mewe \etal\ (2003; see text).  Note that all five stars have
low temperature components of similar value.  $\tau$~Sco, $\xi^1$ CMa,
and HD~63425 all have high temperature components, more so than
$\beta$~CMa.  The hot component of HD~66665 appears consistent
with that of $\beta$~CMa; however, the former has an order of
magnitude fewer counts than the latter.  The error bars are $1\sigma$
values.
\label{fig:counts}}
\end{figure}

As mentioned in the previous section, there is some concern for
HD~66665 that its apparent lack of quite hard emission is an artifact
of its low quality spectrum.  To illustrate Figure~\ref{fig:counts}
plots the two $kT$ values for each source against the total number
of detected X-ray counts.  It seems clear that the {\em failure}
to detect a hot component in $\beta$~CMa, the lone non-magnetic
star in this sample, is not a question of sufficient counts.  Both
HD~63425 and $\xi^1$~CMa have lower total counts in EPIC/PN yet
substantially hotter $kT_2$ values.  The $kT_2$ value for HD~66665
is lowest among the magnetic stars, but suspiciously also has the
lowest total counts.  The fact that the spectrum of HD~66665
is fit by roughly equal amounts of soft and hard emissions is
anomalous among single OB stars, and tantalizingly suggestive that hotter gas may
be present in HD~66665 but was simply not detected.  We feel strongly
that a longer exposure spectrum is needed to determine whether or
not HD~66665 has a hot component, similar to $\tau$~Sco.  Clearly,
HD~63425 does have a hot component similar to $\tau$~Sco.

To place these claims on a more quantitative level, consider the
following analysis of the emission-measure-weighted $kT$ values,
$\langle kT \rangle$, for our sources as compared to our reference
non-magnetic B~star, $\beta$~CMa.
For this purpose we introduce a difference parameter
$\Delta kT$ as

\begin{equation}
\Delta kT = \langle kT \rangle_{\rm star} - \langle kT \rangle_{\rm \beta~CMa} .
\end{equation}

\noindent The error in the difference $\Delta kT$ is given by $\sigma_\Delta$,
with

\begin{equation}
\sigma_\Delta^2 = \sigma_{\rm star}^2 + \sigma_{\rm \beta~CMa}^2.
\end{equation}

\noindent Then the significance of the difference in weighted $kT$
values can be evaluated from the ratio $\Delta kT/\sigma_\Delta$, as 
provided in Table~\ref{tab3}.  The overall harder emission observed
in $\tau$~Sco and HD~63425 as distinct from $\beta$~CMa is found to be significant
at confidence levels of roughly $7\sigma_\Delta$ and $3\sigma_\Delta$, respectively.
For $\xi^1$~CMa the significance is actually less at just 
under $2\sigma_\Delta$.  For HD~66665 the distinction is formally not significant
at all because of the larger error in the determination of $\langle kT \rangle$ for
this star.  

\section{Conclusions}	\label{sec:conc}

The two B stars HD~63425 and HD~66665 have been identified as
analogues to the B~star, $\tau$~Sco, initially owing to similarity
between UV P~Cygni spectral lines.  Given the successful detection
of magnetism in $\tau$~Sco (Donati \etal\ 2006), Petit \etal\ (2011)
reported the search and detection of magnetism in HD~63425 and
HD~66665 that strengthens the physical connection that these three
stars appear to share.

The magnetic detections prompted an investigation of the X-ray
properties of HD~63425 and HD~66665.  The detection of hard emission
with a substantial relative emission measure for HD~63425 appears
to solidify its observational status as a bona fide $\tau$~Sco
analogue by virtue of having (a) peculiar UV wind lines, (b) a
substantial surface magnetic field, and (c) hard X-ray emission
well in excess of values typically seen in single OB stars.

For HD~66665 our analysis does not provide evidence of a substantial
hard component.  However, our observation led to fewer counts than
expected, and we view the experiment as inconclusive.  A longer
exposure is needed to verify whether or not the star possesses
significant plasma of abnormally high temperature as in the case
of $\tau$~Sco.

Given the success in verifying the analogue nature of HD~63425, it
now seems ripe, from an empirical point of view, to suggest that
$\tau$~Sco may be a prototype for a new class of magnetic B~stars.
Exactly how the wind and magnetic field interacts to produce the
observed hard X-ray emissions and pecular UV line morphologies
remain important open questions.  Zeeman Doppler imaging of $\tau$~Sco
has revealed a quite complex magnetic field topology (Donati \etal\
2006).  The star's surface distribution is more complicated than a
simple dipole field like those used in current models of magnetized
stellar winds ud-Doula \& Owocki 2002; Townsend, Owocki, \& ud-Doula
2007).  What can be concluded is that there is a subset of magnetic
B~stars taht share the properties of $\tau$~Sco; and that adopting
$\beta$~CMa as a reference non-magnetic B star, it appears that the
magnetic stars $\tau$~Sco, HD~63425, and $\xi^1$~CMa are all
comparatively hard and X-ray luminous (in terms of $L_X/L_\ast$),
as one might generally expect from models of magnetically channeled
wind flow.  In the future more intensive monitoring of HD~63425
and HD~66665 is needed to discern the detailed magnetic field
geometries of these stars, and more data are needed to confirm
the X-ray nature of HD~66665.

\section*{Acknowledgments}
Based on observations obtained with
\xmm, an ESA science mission with instruments and contributions
directly funded by ESA Member States and NASA.  This research has
made use of NASA's Astrophysics Data System Service and the SIMBAD
database, operated at CDS, Strasbourg, France. We are greatful 
to Alexis Finoguenov for his expert opinion on galaxy cluster X-ray 
emission. Funding for this research has been provided by DLR grant 
50\,OR\,1101 (LMO).

\label{lastpage}

\end{document}